\documentclass
[prd,twocolumn,floatfix,amsmath,nofootinbib,amssymb,floatfix]{revtex4}
\usepackage{graphicx,color,dcolumn,booktabs,bm}
\usepackage{longtable,lscape}
\usepackage{txfonts}
\usepackage{overpic}
\usepackage{amssymb}
\usepackage{array}
\usepackage{indentfirst}
\usepackage{feynmf}   %{feynmp}
\usepackage{slashed}  %for Feynman symbols
\usepackage{cases}
\usepackage{color}
\usepackage{multirow}
\usepackage{epstopdf}
\usepackage{tabularx}
\usepackage[compat=1.1.0]{tikz-feynman}
\usepackage[colorlinks, 
            citecolor=blue,
            anchorcolor=red,
            menucolor=red,
            linkcolor=red,
            filecolor=red,
            runcolor=red,
            urlcolor=blue,
            frenchlinks=red]{hyperref}

\begin{document}
	
	%\begin{CJK}{GBK}{}
\title{Two-body Hidden Charm Decays of $D$ Wave Charmonia}
\author{Xiao-Yu Qi$^{1}$}
\author{Qi Wu$^{2}$}
\author{Xing-Dao Guo$^3$}
\author{Dian-Yong Chen$^{1,4}$\footnote{Corresponding author}}\email{chendy@seu.edu.cn}
\affiliation{
$^1$ School of Physics, Southeast University, Nanjing 210094, China\\
$^2$ Institute of Particle and Nuclear Physics, Henan Normal University, Xinxiang 453007, China\\
$^3$  College of Physics and New Energy, Xuzhou University of Technology, Xuzhou, 221111, China\\
$^4$ Lanzhou Center for Theoretical Physics, Lanzhou University, Lanzhou 730000, China
	}
	\date{\today}

\begin{abstract}
The experimental observations of $\psi_2(3823)$ and $\psi_3(3842)$ make $D$ wave charmonia family abundant. In the present work, we investigate the hidden charm decay processes of spin triplets of the $D$-wave charmonia with the meson loop mechanism. The model parameter $\alpha_\Lambda$ is determined by reproducing the branching fraction of $\psi(3770)\to J/\psi \eta$. With this range of model parameter values, the branching fractions (partial widths) of $\psi(3770) \to \eta_c \omega$, $\psi_2(3823)/\psi_3(3842) \to J/\psi \eta$, $\psi_2(3823)/\psi_3(3842) \to \eta_c \omega$ are estimated. Our estimations find that the partial width of $\psi_2(3823) \to J/\psi \eta$ is $\left(29.64^{+4.01}_{-4.63}\right)\ \mathrm{keV}$, and the partial width ratio of $\psi_2(3823) \to J/\psi \eta$ relative to $\psi_2(3823)\to \gamma \chi_{c1}$ is about $10\%$, which could be tested by further precise measurements from the BESIII, Belle II and LHCb Collaborations.
\end{abstract}
%\pacs{****}
	
\maketitle
%\end{CJK}

\section{introduction}
	
After the observations of $J/\psi$ in 1974 \cite{SLAC-SP-017:1974ind,E598:1974sol}, numerous quark models have made predictions for the charmonium family~\cite{Eichten:1974af, Eichten:1975ag, Lane:1976yh}, prompting experimentalists to search for additional charmonium states and enrich charmonium spectroscopy. The first $D$-wave charmonium, $\psi(3770)$, was reported by SLAC in 1977 in the cross sections for hadron production by $e^+ e^-$ annihilation~\cite{Rapidis:1977cv}. The resonance parameters are measured as $(3772\pm 6)$ MeV and $(28 \pm 5) $ MeV~\cite{Rapidis:1977cv}, which is above the threshold of $D\bar{D}$.

From the perspective of $L-S$ coupling, there exist four $D$ wave states, which are the spin-singlet state $\eta_{c2}$, and the spin-triplet states $\psi_{1}$ (that is, $\psi(3770)$), $\psi_2$ and $\psi_3$. The $\psi_1$ state, characterized by $J^{PC}=1^{--}$, can be easily generated through $e^+ e^-$ annihilation. Conversely, the production mechanisms for $\psi_2$ and $\psi_3$ are more intricate, making their experimental observation more difficult. Almost twenty years after the first observations of $\psi(3770)$, the E705 Collaboration reported a tentative observation of a structure at a mass of 3.836 GeV in the $\pi^ +\pi^- J/\psi$ invariant mass distributions of the process $\pi \mathrm{Li} \to \pi^+ \pi^- J/\psi +X$ in 1994~\cite{E705:1993pry}, which may be consistent with the $\psi(1^3D_2)$ state. In 2013, the Belle Collaboration reported evidence of a new narrow resonance that decays to $\chi_{c1} \gamma$ in $B\to \chi_{c1}\gamma K$ ~\cite{Belle:2013ewt}. This state has a mass of $(3823.1 \pm 1.8 (\mathrm{stat.}) \pm 0.7 (\mathrm{syst.}))$ MeV~\cite{Belle:2013ewt}, which is consistent with the theoretical expectations for the $\psi(1^3D_2)$ charmonium state~\cite{Godfrey:1985xj, Li:2009zu}. Two years later, the BES III Collaboration observed this $D$ wave charmonium state in the $\gamma \chi_{c1}$ invariant mass distributions of the process $e^+ e^- \to \pi^+ \pi^- \gamma \chi_{c1}$ with a statistical significance of 6.2 $\sigma$~\cite{BESIII:2015iqd}. Furthermore, the BES III Collaboration updated their measurements of the cross sections for $e^+ e^- \to \pi^+ \pi^- \psi_2(3823) \to \pi^ +\pi^-\gamma \chi_{c1}$ using a data sample corresponding to an integrated luminosity of 11.3 $\mathrm{fb}^{-1}$ collected at center-of-mass energies from 4.23 to 4.70 GeV, and obtained the most precise measurement of the mass of $\psi_2(3822)$ in 2022~\cite{BESIII:2022yga}. With the same data sample, the BES III Collaboration also reported the first observation of $e^+ e^- \to \pi^0 \pi^0 \psi_2(3823)$~\cite{BESIII:2022cyq}, and the ratio of the average cross sections for the neutral and charged dipion channels was determined to be $0.57 \pm 0.14 \pm 0.05$, which is consistent with the expectations from the isospin symmetry. In Ref.~\cite{LHCb:2020fvo}, the LHCb Collaboration searched for $\psi_2(3823)$ in the $B^+ \to (J/\psi \pi^+ \pi^-) K^+$ decays, and the width of the $\psi_2(3823)$ state was found to be below 5.2 MeV at $90\%$ confidence level. To data, the PDG average of the resonance parameters of $\psi_2(3823)$ are~\cite{ParticleDataGroup:2024cfk}, 
\begin{eqnarray}
	m_{\psi_2} &=& (3823.51 \pm 0.34 ) \ \mathrm{MeV}, \nonumber\\
	\Gamma_{\psi_2} &<&2.9\ \mathrm{MeV},
\end{eqnarray}
respectively.

In 2019, the LHCb Collaboration reported a new narrow charmonium state, $X(3842)$ in the $D\bar{D}$ invariant mass distribution, and the observed mass and width are $(3842.71\pm 0.16 \pm 0.12)$ MeV and $(2.79 \pm 0.51 \pm 0.35)$ MeV, respectively. The observed mass and narrow width suggested the interpretation of $X(3842)$ as the $\psi_3(1^3D_3)$ charmonium state. This state has also been searched in the $D^+D^-$ invariant mass distribution of the $e^+ e^- \to \pi^+ \pi^- D^+ D^-$ process by the BES III Collaboration, and evidence with a significance of $4.2\sigma$ was found in data samples with center-of-mass energies of 4.600 to 4.700 GeV. The PDG average of the resonance parameters of $\psi_3(3843)$ are~\cite{ParticleDataGroup:2024cfk}, 
 \begin{eqnarray}
	m_{\psi_3} &=& (3842.71 \pm 0.20) \ \mathrm{MeV}, \nonumber\\
	\Gamma_{\psi_3} &=& (2.8 \pm 0.6)\ \mathrm{MeV}, 	
 \end{eqnarray}
respectively.

Based on the experimental observations mentioned above, it is noted that $\psi_1$, $\psi_2$ and $\psi_3$ are situated between the $D\bar{D}$ and $D\bar{D}^\ast$ thresholds. $\psi_1$ and $\psi_3$ can couple to $D\bar{D}$ through a $P$ and $F$ wave, respectively, while coupling between $\psi_2$ and $D\bar{D}$ is prohibited due to conservations of $J^P$ quantum numbers. Thus, the radiative and hidden-charm strong decay modes should be dominant for $\psi_2(3823)$. Moreover, for $\psi(3770)$, the BES Collaboration's analyses have revealed that it possesses a substantial non-$D\bar{D}$ branching fraction, which was measured to be $ (16.4\pm 7.3\pm 4.2)\%$~\cite{BES:2006fpf}, $(16.1 \pm 1.6 \pm 5.7)\%$~\cite{BES:2006dso}, and $ (15.1 \pm 5.6 \pm 1.8)\%$~\cite{BES:2008vad}, with variations attributed to different analysis techniques and independent data samples. 

Hidden-charm decay processes, being a category of non$-D\bar{D}$ channel, play a vital role in elucidating the surprisingly large branching fraction of non-$D\bar{D}$ observed in $\psi(3770)$. The most recent measurement of the process $\psi(3770) \to J/\psi \eta$ revealed a branching fraction of $(8.7\pm  1.0 \pm 0.8) \times 10^{-4}$, yet it remains insufficient to fully comprehend the substantial non-$D\bar{D}$ branching fraction of $\psi(3770)$. Therefore, the exploration of additional non-$D\bar{D}$ channels of $\psi(3770)$ is of great interest. Furthermore, $\psi_2(3823)$ cannot decay into $D\bar{D}$, thus the hidden charm decay channels are even more important for understanding the decay properties of $\psi_2(3823)$. Although $\psi_3(3842)$ can decay to $D\bar{D}$, this process only occurs through a $F-$wave, which should be suppressed. Thus, the hidden charm decay processes also remain important for $\psi_3(3842)$. Thus, in the present work, we perform a systematic investigation of the two-body hidden charm decay processes of the $D$-wave charmonia, aiming to provide further insight into the decay properties of these charmonia.

It is worth mentioning that to understand the unexpected large non-$D\bar{D}$ branching fraction of $\psi(3770)$, the authors in Refs.~\cite{Zhang:2009kr, Liu:2009dr} employed the meson loops to solve the puzzle of the excessive non-$D\bar{D}$ component of the inclusive $\psi(3770)$ decay. Such meson loops could be considered as phenomenological descriptions of the long-distant contributions to the OZI suppressed decay process, which have been applied to various channels~\cite{Guo:2009wr,Guo:2010ak,Li:2007xr,Liu:2009dr,Liu:2009iw,He:2006is,Chen:2009ah,Chen:2010re, Chen:2011qx,Chen:2011jp,Chen:2012nva,Chen:2013cpa,Chen:2013yxa}. Thus, in the present work, we extend the meson loop mechanism to investigate the hidden charm decays of $D$-wave charmonia.

This work is organized as follows. After the introduction, we present our estimations of the hidden charm decays of the $D-$wave charmonia through the meson loop mechanism. While Section \ref{Sec:Num} presents the numerical results and discussions and a brief summary is provided in Section \ref{Sec:Sum}.

\begin{figure}[t]
	\begin{tabular}{ccc}
		\centering
		\includegraphics[width=4.2cm]{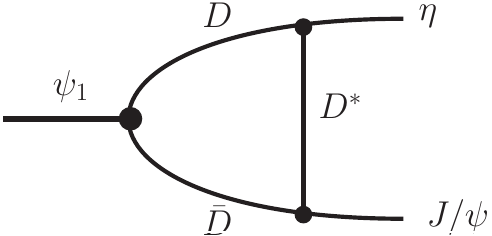}&
		\includegraphics[width=4.2cm]{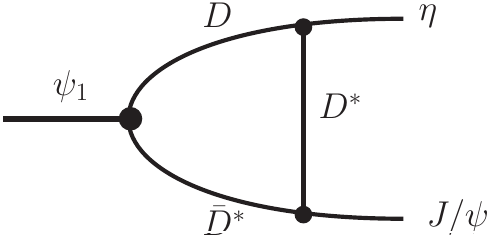}\\ \\
		$(a)$ & $(b)$ \\ \\
		\includegraphics[width=4.2cm]{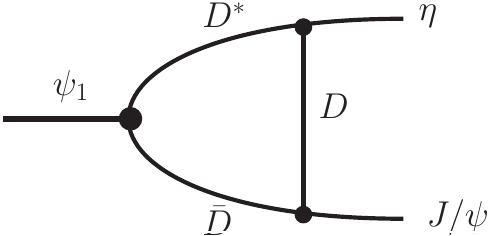}&
		\includegraphics[width=4.2cm]{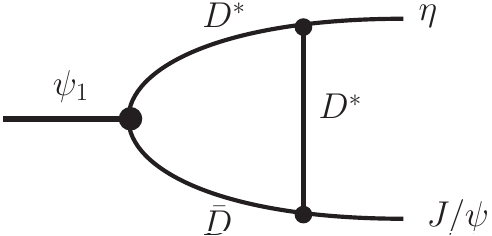}\\ \\
	    $(c)$ & $(d)$\\ \\
		\includegraphics[width=4.2cm]{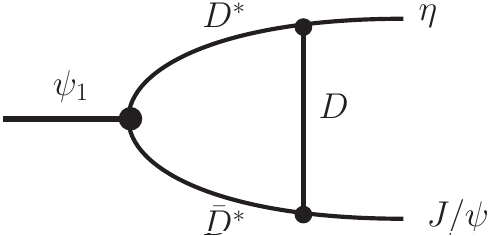}&
		\includegraphics[width=4.2cm]{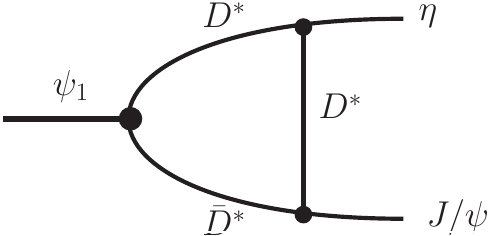}\\ \\
	    $(e)$ & $(f)$ \\
	\end{tabular}
	\caption{Diagrams contributing to $\psi(3770)\to J/\psi\eta$ at the hadron level.\label{Fig:Tri1}}
\end{figure}

\begin{figure}[t]
	\begin{tabular}{ccc}
		\centering
		\includegraphics[width=4.2cm]{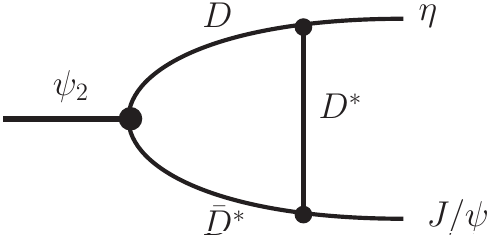}&
		\includegraphics[width=4.2cm]{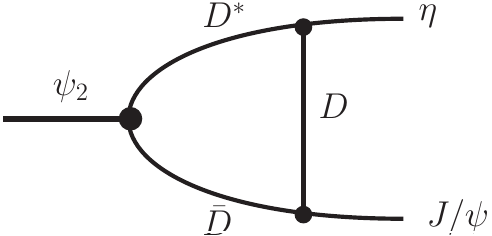} \\ \\
		$(g)$ & $(h)$ \\ \\
		\includegraphics[width=4.2cm]{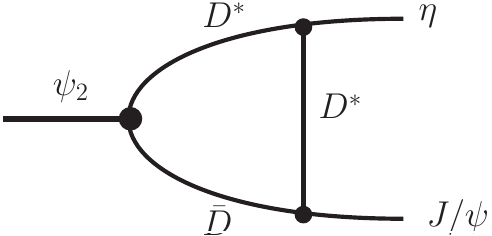}&
		\includegraphics[width=4.2cm]{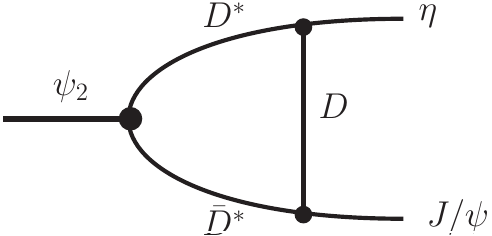} \\ \\
	    $(i)$ & $(j)$ \\ \\
		\includegraphics[width=4.2cm]{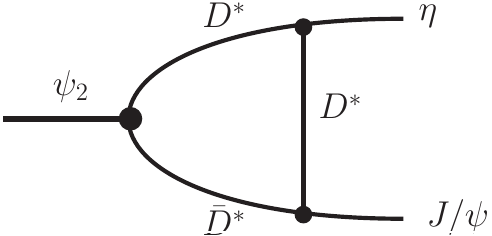}\\ \\
		$(k)$  
	\end{tabular}
	\caption{The same as Fig.~\ref{Fig:Tri1} but for $\psi_2(3823)\to J/\psi \eta$. \label{Fig:Tri2}}
\end{figure}

\begin{figure}[t]
	\begin{tabular}{ccc}
		\centering
		\includegraphics[width=4.2cm]{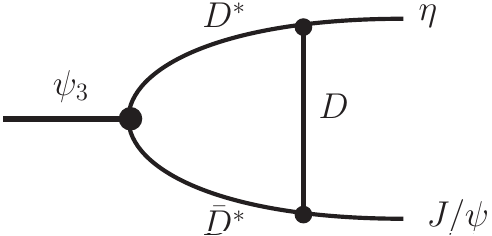}&
		\includegraphics[width=4.2cm]{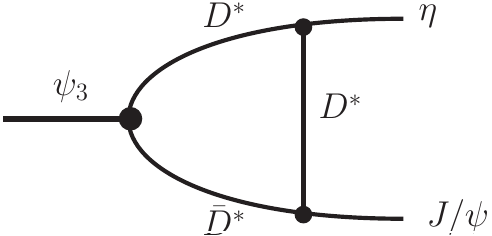}&
		\\ \\
		$(l)$ & $(m)$  \\
		\\		
	\end{tabular}
	\caption{The same as Fig.~\ref{Fig:Tri1} but for $\psi_3(3842)\to J/\psi \eta$. \label{Fig:Tri3}}
\end{figure}

\section{THE HIDDEN CHARM DECAYS of $\psi(3770),\psi_2(3823) $ and $\psi_3(3842)$}
\label{sec:Sec2}

Considering the conservations of the $J^{PC}$ quantum numbers and the kinematic limit, the possible two-body hidden charm decay channels of the $D-$wave charmonia are $J/\psi \eta$ and $\eta_c \omega$. As indicated in Refs.~\cite{Zhang:2009kr, Liu:2009dr, Li:2013zcr}, the meson loop mechanism, as a phenomenological description of the long-distant contributions to the OZI suppressed process, essentially contributes to the non-$D\bar{D}$ decay channels of $\psi(3770)$. Considering the fact that the masses of the $D$ wave ground charmonia states are far below the thresholds of a $S$-wave and a $P$-wave charmed mesons, we only consider the meson loops constituted by $S$-wave charmed mesons in the present estimations. The relevant meson loop diagrams for $\psi(3770)/\psi_2(3823)/\psi_3(3842)\to J/\psi \eta$ are presented in Figs.~\ref{Fig:Tri1}-\ref{Fig:Tri3}. The diagrams contributing to $\psi(3770)/\psi_2(3823)/\psi_3(3842)\to  \eta_c \omega$ are collected in the Appendix \ref{Sec:AppB}.

In the present estimations, we evaluate the diagrams in Figs.~\ref{Fig:Tri1}-\ref{Fig:Tri3} at the hadron level with an effective Lagrangian, which is constructed based on the heavy quark limit and chiral symmetry. In the heavy quark limit, the total angular momentum of the heavy-light meson can be decoupled as $\vec{J}= \vec{S}_Q +\vec{S}_\ell$ with $\vec{S}_Q$ and $\vec{S}_\ell$ being the spin of the heavy quark and the light degree of freedom, respectively. While $\vec{S}_\ell$ is the sum of the spin of light antiquark and the orbital angular momentum, i.e., $\vec{S}_\ell=\vec{S}_{\bar{q}}+\ell$. In the heavy quark limit, the heavy quark is infinitely heavy, which leads to the spin of the heavy quark $\vec{S}_Q$ and the total angular momentum of the light degree of freedom $\vec{S}_\ell$ being separately conserved by strong interactions. In the heavy quark effective theory, the states have the same $\vec{S}_\ell$ are degenerated. For example, for the $S$-wave charmed or charmed strange mesons, they are degenerated and could be expressed by the superfield $H_1$, which is~\cite{Mannel:1997ky, Brambilla:2004jw, Colangelo:2003sa},
\begin{eqnarray}
H_1&=&\frac{1+v\!\!\!\slash}{2}\left[\mathcal{D}^{\ast\mu}\gamma_\mu-\mathcal{D}\gamma_5\right],
\end{eqnarray}
with $\mathcal{D}^{(\ast)}=(D^{(\ast)0}, D^{(\ast) +}, D_s^{(\ast)+})$, and the corresponding anti-charmed or anti-charmed strange mesons is,
 \begin{eqnarray}
H_2&=&\left[\bar{\mathcal{D}}^{\ast\mu}\gamma_\mu-\bar{\mathcal{D}}\gamma_5\right]\frac{1-v\!\!\!\slash}{2}.
\end{eqnarray}

Similarly, for the heavy quarkonium, the degeneracy is expected under rotations of the two heavy-quark spins, which allows us to build up heavy quarkonia multiplets for a certain relative angular momentum $\ell$. The $S-$wave charmonia with $\ell =0$ could be expressed by the $4\times 4$ matrix, which reads~\cite{Mannel:1997ky, Brambilla:2004jw, Colangelo:2003sa}, 
\begin{eqnarray}
	\mathcal{R}&=&\frac{1+v\!\!\!\slash}{2}\left[\psi^{\mu}\gamma_\mu-\eta_{c}\gamma_5\right]\frac{1-v\!\!\!\slash}{2}.
\end{eqnarray}
For $\ell=2$, four charmonia, which are degenerate in the heavy quark limit,  can be build, and the corresponding spin multiplet reads~\cite{Mannel:1997ky, Brambilla:2004jw, Colangelo:2003sa},
\begin{eqnarray}
	\mathcal{J}^{\mu\lambda}&=&\frac{1+v\!\!\!\slash}{2}\left[\psi^{\mu\alpha\lambda}_3 \gamma_\alpha+\frac{1}{\sqrt{6}}\Big(\varepsilon^{\mu\alpha\beta\rho}v_\alpha \gamma_\beta \psi^\lambda_{2\rho}+\varepsilon^{\lambda\alpha\beta\rho}v_\alpha \gamma_\beta \psi^\mu_{2\rho}\Big)\right.\nonumber\\
		&&+\frac{\sqrt{15}}{10}\Big((\gamma^\mu-v^\mu)\psi^\lambda_1+(\gamma^\lambda-v^\lambda)\psi^\mu_1\Big)\nonumber\\
		&&-\left.\frac{1}{\sqrt{15}}\Big(g^{\mu\lambda}-v^\mu v^\lambda\Big)\gamma_\alpha \psi^\alpha_1+\eta^{\mu\lambda}_{c2}\gamma_5\right]\frac{1-v\!\!\!\slash}{2}.
\end{eqnarray}

The general Lagrangians describing the $S-$ and $D-$wave charmonia to a pair of charmed and anti-charmed mesons read~\cite{Mannel:1997ky, Brambilla:2004jw, Colangelo:2003sa},
\begin{eqnarray}
	\mathcal{L_S}=ig_1 \mathrm{Tr}\Big[\mathcal{R}\bar{H}_2 {\stackrel{\leftrightarrow}{\partial_\mu}}\gamma_\lambda \bar{H}_1\Big]+\mathrm{H.c.},\nonumber\\
		\mathcal{L_D}=g_2 \mathrm{Tr}\Big[\mathcal{J}^{\mu\lambda}\bar{H}_2 {\stackrel{\leftrightarrow}{\partial_\mu}}\gamma_\lambda \bar{H}_1\Big]+\mathrm{H.c.},
		\label{Eq:LagT}
	\end{eqnarray}
with $\bar{H}_{1,2} =\gamma^0 H_{1,2}^\dagger \gamma^0$. By further expanding the Lagrangian mentioned above, we can derive the relevant effective Lagrangians involved in the present estimations, which are,
\begin{eqnarray}
		\mathcal{L}_{\psi \mathcal{D}^{(*)}\mathcal{D}^{(*)}}&=&-ig_{\psi \mathcal{D} \mathcal{D}}\psi_{\mu}\Big(\partial^{\mu}\mathcal{\mathcal{D}\mathcal{D}}^{\dagger}-\mathcal{D}\partial^{\mu}\mathcal{D}^{\dagger}\Big)\nonumber \\
		&&+g_{\psi \mathcal{D}^* \mathcal{D}}\varepsilon^{\mu\nu\alpha\beta}\partial_{\mu}\psi_{\nu}\Big(\mathcal{D}_{\alpha}^{*}\stackrel{\leftrightarrow}{\partial}_{\beta}\mathcal{D}^{\dagger}-\mathcal{D} \stackrel{\leftrightarrow}{\partial_{\beta}} \mathcal{D}_{\alpha}^{*\dagger}\Big) \nonumber\\
		&&+ig_{\psi \mathcal{D}^* \mathcal{D}^*}\psi^\mu\Big(\mathcal{D}_\nu^*\stackrel{\leftrightarrow}{\partial^\nu} \mathcal{D}_\mu^{*\dagger}+\mathcal{D}_\mu\stackrel{\leftrightarrow}{\partial^\nu} \mathcal{D}_\nu^{*\dagger}\nonumber\\
		&&-\mathcal{D}_\nu \stackrel{\leftrightarrow}{\partial_\mu} \mathcal{D}^{*\dagger\nu}\Big),\nonumber\\
	\mathcal{L}_{\eta_c\mathcal{D}^*\mathcal{D}^{(*)}}&=&-ig_{\eta_c \mathcal{D}^*\mathcal{D}}\eta_c \Big(\mathcal{D}\stackrel{\leftrightarrow}{\partial_\mu}\mathcal{D}^{*\dagger\mu}+\mathcal{D}^{*\mu}\stackrel{\leftrightarrow}{\partial_\mu}\mathcal{D}^\dagger\Big)\nonumber\\
		&&-g_{\eta_c \mathcal{D}^*\mathcal{D}^*}\varepsilon^{\mu\nu\alpha\beta}\partial_{\mu}\eta\mathcal{D}^{*}_{\nu}\stackrel{\leftrightarrow}{\partial_\alpha}\mathcal{D}^{*\dagger}_\beta ,\nonumber\\
	\mathcal{L}_{\psi_J \mathcal{D}^{(*)}\mathcal{D}^{(*)}}&=&g_{\psi_1 \mathcal{D} \mathcal{D}}\psi^\mu_1 \Big(\mathcal{D} \partial_\mu \mathcal{D}^\dagger-\mathcal{D}^\dagger\partial_\mu\mathcal{D}\Big)\nonumber\\
		&&+g_{\psi_1 \mathcal{D} \mathcal{D}^\ast}\varepsilon^{\mu\nu\alpha\beta}\Big(\mathcal{D}{\stackrel{\leftrightarrow}{\partial_\mu}}\mathcal{D}^{\ast\dagger}_\beta-\mathcal{D}^{\ast}_\beta{\stackrel{\leftrightarrow}{\partial_\mu}}\mathcal{D}^\dagger\Big)\partial_\nu \psi_{1\alpha}\nonumber\\
		&&+g_{\psi_1 \mathcal{D}^\ast \mathcal{D}^\ast}\left[-4\Big(\psi^\mu_1 \mathcal{D}^{\ast\nu\dagger}\partial_\mu \mathcal{D}^{\ast}_\nu-\psi^\mu_1 \mathcal{D}^{\ast}_\nu \partial_\mu \mathcal{D}^{\ast\nu\dagger}\Big)\right.\nonumber\\
		&&+\left.\psi^\mu_1 \mathcal{D}^{\ast\nu\dagger}\partial_\nu \mathcal{D}^{\ast}_\mu-\psi^\mu_1 \mathcal{D}^{\ast\nu}\partial_\nu \mathcal{D}^{\ast\dagger}_\mu\right]\nonumber\\
		&&+ig_{\psi_2 \mathcal{D} \mathcal{D}^\ast}\psi^{\mu\nu}_2\Big(\mathcal{D}{\stackrel{\leftrightarrow}{\partial_\nu}}\mathcal{D}^{\ast\dagger}_\mu-\mathcal{D}^{\ast}_\mu{\stackrel{\leftrightarrow}{\partial_\nu}}\mathcal{D}^\dagger\Big)\nonumber\\
		&&+ig_{\psi_2 \mathcal{D}^\ast \mathcal{D}^\ast}\varepsilon_{\alpha\beta\mu\nu}\Big(\mathcal{D}^{\ast\nu}{\stackrel{\leftrightarrow}{\partial^\beta}}\mathcal{D}^{\ast\dagger}_\lambda-\mathcal{D}^{\ast\nu\dagger}{\stackrel{\leftrightarrow}{\partial^\beta}}\mathcal{D}^{\ast}_\lambda\Big)\partial^\mu \psi^{\alpha\lambda}_2\nonumber\\
		&&+g_{\psi_3 \mathcal{D}^\ast \mathcal{D}^\ast}\psi^{\mu\nu\alpha}_3 \Big(\mathcal{D}^{\ast}_\alpha{\stackrel{\leftrightarrow}{\partial_\mu}}\mathcal{D}^{\ast\dagger}_\nu+\mathcal{D}^{\ast}_\nu{\stackrel{\leftrightarrow}{\partial_\mu}}\mathcal{D}^{\ast\dagger}_\alpha\Big).
	\end{eqnarray}
	
Taking into account the heavy quark limit and the chiral symmetry as well, one can construct the  the effective interaction involving light pseudoscalar (vector) mesons and charm mesons, which are ~\cite{Falk:1992cx,Chen:2014sra,Yan:1992gz,Cheng:1992xi,Wise:1992hn},
	\begin{eqnarray}
		\mathcal{L}_{\mathcal{D}^{(*)} \mathcal{D}^{(*)}\mathcal{P}}&=&-ig_{\mathcal{D}^\ast \mathcal{D} \mathcal{P}}\Big(\mathcal{D}^{i\dagger} \partial^\mu \mathcal{P}_{ij}\mathcal{D}^{\ast j}_\mu-\mathcal{D}^{\ast i\dagger}_\mu \partial^\mu \mathcal{P}_{ij}\mathcal{D}^{j}\Big)\nonumber\\
		&&+\frac{1}{2}g_{\mathcal{D}^\ast \mathcal{D}^\ast \mathcal{P}}\varepsilon_{\mu\nu\alpha\beta}\mathcal{D}^{\ast \mu\dagger}_i \partial^\nu \mathcal{P}_{ij}{\stackrel{\leftrightarrow}{\partial^\alpha}}\mathcal{D}^{\ast \beta}_j,\nonumber\\
		\mathcal{L}_{\mathcal{D}^{(*)} \mathcal{D}^{(*)}\mathcal{V}}&=&-i g_{\mathcal{D} \mathcal{D} \mathcal{V}} \mathcal{D}_{i}^{\dagger} \stackrel{\leftrightarrow}\partial_{\mu} \mathcal{D}^{j}\left(\mathcal{V}^{\mu}\right)_{j}^{i}-2 f_{\mathcal{D}^{*} \mathcal{D} \mathcal{V}} \varepsilon^{\mu \nu \alpha \beta}\nonumber \\
		&& \times\left(\partial_{\mu} \mathcal{V}_{\nu}\right)_{j}^{i}\left(\mathcal{D}_{i}^{\dagger}\stackrel{\leftrightarrow} \partial_{\alpha} \mathcal{D}_{\beta}^{* j}-\mathcal{D}_{\beta i}^{* \dagger} \stackrel{\leftrightarrow}{\partial}_{\alpha} \mathcal{D}^{j}\right)\nonumber \\
		&&+i g_{\mathcal{D}^{*} \mathcal{D}^{*} \mathcal{V}} \mathcal{D}_{i}^{* \dagger \nu} \stackrel{\leftrightarrow}{\partial_{\mu}} \mathcal{D}_{\nu}^{* j}\left(\mathcal{V}^{\mu}\right)_{j}^{i}\nonumber \\
		&&+4 i f_{\mathcal{D}^{*} \mathcal{D}^{*} \mathcal{V}} \mathcal{D}_{i}^{* \dagger \mu}\left(\partial_{\mu} \mathcal{V}^{\nu}-\partial^{\nu} \mathcal{V}_{\mu}\right)_{j}^{i} \mathcal{D}_{\nu}^{* j},
    \end{eqnarray}
with $\mathcal{P}$ and $\mathcal{V}$ being the pseudoscalar and vector mesons, respectively, and their matrices forms are~\cite{Casalbuoni:1996pg,Casalbuoni:1992gi,Casalbuoni:1992dx},
	\begin{eqnarray}
 \mathcal{P}&=&\left(\begin{array}{ccc}
			\frac{\pi^{0}}{\sqrt{2}}+\alpha \eta+\beta \eta \prime & \pi^{+} & K^{+} \\
			\pi^{-} & -\frac{\pi^{0}}{\sqrt{2}}+\alpha \eta+\beta \eta \prime & K^{0} \\
			K^{-} & \bar{K}^{0} & \gamma \eta+\delta \eta^{\prime}
		\end{array}\right), \nonumber
	\end{eqnarray}
	\begin{eqnarray} 
		\mathcal{V} &=& \left(\begin{array}{ccc}
			\frac{1}{\sqrt{2}}(\rho^{0}+\omega ) & \rho^{+} & K^{*+} \\
			\rho^{-} & \frac{1}{\sqrt{2}}(-\rho^{0}+\omega) & K^{* 0} \\
			K^{*-} & \bar{K}^{* 0} & \phi
		\end{array}\right),
	\end{eqnarray}
where the parameters $\alpha$, $\beta$, $\gamma$ and~$\delta$ are the parameters associate to the mixing angle $\theta$, which are defined as~\cite{Gilman:1987ax}, 
\begin{eqnarray}
\alpha=\frac{\mathrm{cos}\theta-\sqrt{2}\mathrm{sin}\theta}{\sqrt{6}}&, \quad& \beta=\frac{\mathrm{sin}\theta+\sqrt{2}\mathrm{cos}\theta}{\sqrt{6}}, \nonumber\\
\gamma=\frac{-2\mathrm{cos}\theta-\sqrt{2}\mathrm{sin}\theta}{\sqrt{6}}&, \quad& \delta=\frac{-2\mathrm{sin}\theta+\sqrt{2}\mathrm{cos}\theta}{\sqrt{6}}. \quad 
\end{eqnarray}
where $\theta$ is the mixing angle, ranging from  $-10^\circ$ to $-20^\circ$\cite{Feldmann:1999uf, Mo:2024jjd}. In our current calculations, we set the mixing angle $\theta=19.1^{\circ}$ \cite{MARK-III:1988crp, DM2:1988bfq}. %%%%	

Utilizing the effective Lagrangians mentioned above, we can obtain the amplitudes of the considered hidden charm decay processes of the $D$-wave charmonia. Taking  $\psi(3770)(p_0) \to [D^{(\ast)}(p_1) \bar{D}^{(\ast)}(p_2)] D^{(\ast)}(q) \to \eta(p_3) J/\psi(p_4)$ as an example, the amplitudes corresponding to the diagrams in Fig.~\ref{Fig:Tri1}-$(a)$-$(f)$ are,
\begin{eqnarray}
	\mathcal{M}_{a}&=&i^3 \int\frac{d^4 q}{(2\pi)^4}\left[
		ig_{\psi_1 D D} \epsilon^{\mu}(p)\left(p_{2 \mu}- p_{1 \mu}\right)\right]
		\nonumber\\
		&&\left[-g_{\eta D^{*} D} p_{3\chi }\right]\left[g_{J / \psi D^{*} D} \varepsilon _{\kappa\tau\phi\xi}\left(p^{\xi}_{2}-q^{\xi}\right)  p_{4}^{\kappa } \epsilon^{\tau}(p_{4}) \right]
		\nonumber\\
		&&\times\frac{1}{p^2_1-m^2_1}\frac{1}{p^2_2-m^2_2}\frac{-g_{\chi\phi}+q^{\chi} q^{\phi} /m^2_q}{q^2-m^2_q}\mathcal{F}^2(m_{q}^2, q^2),\nonumber\\
%%
%%		%2
		\mathcal{M}_{b}&=&i^3 \int\frac{d^4 q}{(2\pi)^4}\left[
		g_{\psi_1 D^{*} D} \varepsilon_{\mu \nu \alpha \beta}\left(p^{\mu}_{2}-p^{\mu}_{1}\right)  p^{v}\epsilon^{\alpha}\left(p\right)\right]\nonumber\\
		&&\left[-g_{\eta D^{*} D} p_{3\chi }\right]\left[-g_{J / \psi D^{*} D^{*}} \epsilon^{\kappa}\left(p_{4}\right)\left(\left(q^{\xi}-p^{\xi}_{2}\right) g_{\kappa \phi}\right.\right.\nonumber\\
		&&+\left.\left.(q^{\phi}-p^{\phi}_{2})g_{\kappa \xi }-(q_{\kappa}-p_{2\kappa})g_{\xi \phi}\right)\right] \frac{1}{p^2_1-m^2_1}\nonumber\\
		&&\times\frac{-g_{\beta\phi}+p_2^{\beta} p_2^{\phi} /m^2_2}{p^2_2-m^2_2}\frac{-g_{\chi\xi}+q^{\chi} q^{\xi} /m^2_q}{q^2-m^2_q}\mathcal{F}^2(m_{q}^2, q^2),\nonumber\\
		%3
		\mathcal{M}_{c}&=&i^3 \int\frac{d^4 q}{(2\pi)^4}\left[
		g_{\psi_{1} D^{*} D} \varepsilon_{\mu \nu \alpha \beta} \left(p^{\mu}_{1}-p^{\mu}_{2}\right)  p^{\nu} \epsilon^{\alpha}(p)\right]\nonumber\\
		&&\left[g_{\eta D^{*} D} p_{3\chi }\right]\left[g_{J / \psi D D} \epsilon^{\kappa}\left(p_{4}\right) \left(p_{2\kappa}-q_{\kappa}\right)\right]
		\nonumber\\
		&&\times\frac{-g_{\beta\chi}+p_1^{\beta} p_1^{\chi} /m^2_1}{p^2_1-m^2_1}\frac{1}{p^2_2-m^2_2}\frac{1}{q^2-m^2_q}\mathcal{F}^2(m_{q}^2, q^2),\nonumber\\
		%4
		\mathcal{M}_{d}&=&i^3 \int\frac{d^4 q}{(2\pi)^4}\left[
		g_{\psi_{1} D^{*} D} \varepsilon_{\mu \nu \alpha \beta} \left(p^{\mu}_{1}-p^{\mu}_{2}\right)  p^{\nu} \epsilon^{\alpha}(p)\right]\nonumber\\
		&&\left[\frac{1}{2} g_{\eta D^{*} D^{*} } \varepsilon_{\chi\lambda\sigma\rho}  p_{3}^{\lambda} \left(p^{\sigma}_{1}+q^{\sigma}\right) \right]\left[g_{J / \psi D^{*} D} \varepsilon _{\kappa\tau\phi\xi}\right. \nonumber\\
		&&\left.\left(p^{\xi}_{2}-q^{\xi}\right)  p_{4}^{\kappa } \epsilon^{\tau}\left(p_{4}\right)\right]\frac{1}{p^2_2-m^2_2}
		\nonumber\\
		&&\times\frac{-g_{\beta\rho}+p_1^{\beta} p_1^{\rho} /m^2_1}{p^2_1-m^2_1}\frac{-g_{\chi\phi}+q^{\chi} q^{\phi} /m^2_q}{q^2-m^2_q}\mathcal{F}^2(m_{q}^2, q^2),\nonumber
\end{eqnarray}
\begin{eqnarray}
		%5
		\mathcal{M}_{e}&=&i^3 \int\frac{d^4 q}{(2\pi)^4}\left[ig_{\psi_{1} D^{*} D^{*}}\epsilon^{\mu}(p)\left(-4 ( p_{1\mu}-p_{2\mu})  g^{\alpha \beta}\right.\right.\nonumber\\
		&&\left.\left.+p^{\alpha}_{1}g_{\mu}^{\beta}-p^{\beta}_{2}g^{\alpha}_{\mu}\right) \right] \left[g_{\eta D^{*} D}p_{3\chi }\right]\nonumber\\
		&&\left[
		g_{J / \psi D^{*} D} \varepsilon _{\kappa\tau\phi\xi}\left(q^{\xi}-p^{\xi}_{2}\right) p_{4}^{\kappa } \epsilon^{\tau}(p_{4})\right]
		\nonumber\\
		&&\times\frac{-g_{\beta\chi}+p_1^{\beta} p_1^{\chi} /m^2_1}{p^2_1-m^2_1}\frac{-g_{\alpha\phi}+p_2^{\alpha} p_2^{\phi} /m^2_2}{p^2_2-m^2_2}\nonumber\\
		&&\times\frac{1}{q^2-m^2_q}\mathcal{F}^2(m_{q}^2, q^2),\nonumber\\
%%
		%6
		\mathcal{M}_{f}&=&i^3 \int\frac{d^4 q}{(2\pi)^4}\left[ig_{\psi_{1} D^{*} D^{*}}\epsilon^{\mu}(p)\left(-4 ( p_{1\mu}-p_{2\mu})  g^{\alpha \beta}\right.\right.\nonumber\\
		&&\left.\left.+p^{\alpha}_{1}g_{\mu}^{\beta}-p^{\beta}_{2}g^{\alpha}_{\mu}\right) \right] \Big[\frac{1}{2} g_{\eta D^{*} D^{*} } \varepsilon_{\chi\lambda\sigma\rho}  p_{3}^{\lambda} \nonumber\\
		&&\left. (p^{\sigma}_{1}+q^{\sigma}) \right]\left[-g_{J / \psi D^{*} D^{*}} \epsilon^{\kappa}(p_{4})\left((q_{\xi}-p_{2\xi}) g_{\kappa \phi} \right.\right.\nonumber\\
		&&\left.\left.+\left(q_{\phi}-p_{2\phi}\right)g_{\kappa \xi } -\left(q_{\kappa}-p_{2\kappa}\right)g_{ \phi\xi }\right)\right]
		\nonumber\\
		&&\times\frac{-g_{\beta\rho}+p_1^{\beta} p_1^{\rho} /m^2_1}{p^2_1-m^2_1}\frac{-g_{\alpha\phi}+p_2^{\alpha} p_2^{\phi} /m^2_2}{p^2_2-m^2_2}\nonumber\\
		&&\times\frac{-g^{\chi\xi}+q^{\chi} q^{\xi} /m^2_q}{q^2-m^2_q}\mathcal{F}^2(m_{q}^2, q^2).
	\end{eqnarray}
	
In the amplitudes mentioned above, a form factor $\mathcal{F}(m_q^2,q^2)$ is introduced to ensure the convergence of the amplitude in the ultraviolet region and to account for off-shell effects. In the present work, we take the form factor  $\mathcal{F}(m_q^2,q^2)$ in the monopole form, which is, 
	\begin{eqnarray}
		\mathcal{F}\left(m_q^2, q^{2}\right)=\frac{m_q^{2}-\Lambda^{2}}{q^{2}-\Lambda^{2}},
	\end{eqnarray}
where the parameter $\Lambda$ can be reparameterized as $\Lambda=m+\alpha_\Lambda \Lambda_{\mathrm QCD}$ with $\Lambda_{\mathrm QCD}=220$ MeV. Generally, the model parameter $\alpha_\Lambda$ should be of the order of unity, but its concrete value cannot be determined from first principles. A more practical approach when addressing particular processes is to ascertain the value of $\alpha_\Lambda$ by comparing experimental measurement results with theoretical predictions~\cite{Tornqvist:1993vu,Tornqvist:1993ng,Locher:1993cc,Li:1996yn}.
	
With the above amplitudes, one can obtain the amplitude for $\psi(3770) \to J/\psi \eta $, which is,
\begin{eqnarray}
\mathcal{M}_{\psi_1 \to J/\psi \eta} = \mathcal{M}_a +\mathcal{M}_b +\mathcal{M}_c+\mathcal{M}_d+\mathcal{M}_e+\mathcal{M}_f,
\end{eqnarray} 
and the decay width of $\psi(3770) \to J/\psi \eta$ could be evaluated as,
	\begin{eqnarray}
		\Gamma_{\psi_1 \to J/\psi \eta}=\frac{1}{3} \frac{1}{8 \pi} \frac{|\vec{p}|}{m_{\psi_1}^{2}} \overline{\left|\mathcal{M}_{\psi \to J/\psi \eta}\right|^{2}},
	\end{eqnarray}
where the factor $1/3$ comes from the average the spin of $\psi_1(3770)$, and the overline indicates the sum over the spin of the involved particles. Similarly, we can obtain the amplitudes corresponding to the diagrams in Figs.~\ref{Fig:Tri2}-\ref{Fig:Tri3}, which are collected in Appendix \ref{Sec:AppA}.

\section{Numerical Results and discussion}
\label{Sec:Num}
\subsection{Coupling Constants }
Before estimating the partial widths of the considered processes, the coupling constants in the effective Lagrangians should be further clarified. In the heavy quark limit, the coupling constants relevant  to $S-$wave charmonia and charmed mesons could be related to a gauge coupling $g_1$ by~\cite{Falk:1992cx,Chen:2014sra,Yan:1992gz,Cheng:1992xi,Wise:1992hn},
\begin{eqnarray}
	g_{\psi D D} &=&2 g_{1} \sqrt{m_{\psi}} m_{D}\  ,\nonumber\\
	g_{\psi D^{*} D} &=&2 g_{1} \sqrt{m_{D} m_{D^{*}} / m_{\psi}}\ ,\nonumber \\
	g_{\psi D^{*} D^{*}} &=&2 g_{1} \sqrt{m_{\psi}} m_{D^{*}}\ ,\nonumber\\
	g_{\eta_{c} D^{*} D}&=&2 g_{1} \sqrt{m_{\eta_{c}} m_{D} m_{D^{*}}}\ , \nonumber\\		
	g_{\eta_{c} D^{*} D^{*}}&=&2 g_{1} m_{D^{*}} / \sqrt{m_{\eta_{c}}}\ ,
	\end{eqnarray}
where the gauge coupling is defined as $g_1= \sqrt{m_{\psi}}/\left(2m_Df_\psi \right)$ with  $f_\psi=416$ MeV to be the decay constant of $J/\psi$, which can be estimated from the dilepton decay width of $J/\psi$~\cite{Colangelo:2003sa,ParticleDataGroup:2024cfk}.

For the $D$-wave charmonia, the relevant coupling constants could related to a gauge coupling constant $g_2$ by~\cite{Qi:2023kwc},
\begin{eqnarray}
	g_{\psi_1 DD}&=&-2g_2 \frac{\sqrt{15}}{3}\sqrt{m_{\psi_1}m_D m_D}\ ,\nonumber\\
	g_{\psi_1 DD^\ast}&=&g_2 \frac{\sqrt{15}}{3}\sqrt{m_{D}m_{D^\ast}/m_{\psi_1}}\ ,\nonumber\\
	g_{\psi_1 D^\ast D^\ast}&=&-g_2 \frac{\sqrt{15}}{15}\sqrt{m_{\psi_1}m_{D^\ast} m_{D^\ast}}\ ,\nonumber\\
		g_{\psi_2 D D^\ast}&=&2g_2 \sqrt{\frac{3}{2}}\sqrt{m_{\psi_2}m_D m_{D^\ast}}\ ,\nonumber\\
	g_{\psi_2 D^\ast D^\ast}&=&-2g_2 \sqrt{\frac{1}{6}}\sqrt{m_{D^\ast}m_{D^\ast}/m_{\psi_2}}\ ,\nonumber\\
	g_{\psi_3 D^\ast D^\ast}&=&2g_2 \sqrt{m_{\psi_3}m_{D^\ast} m_{D^\ast}}\ ,
\end{eqnarray}
with  $g_2=1.39$, which is obtained by the partial width of $\psi(3770) \to D\bar{D}$~\cite{ParticleDataGroup:2024cfk}. 

Considering the heavy quark limit and chiral symmetry, one can obtain the coupling constants related to the light pseudoscalar and vector mesons, which are~\cite{Falk:1992cx,Chen:2014sra,Yan:1992gz,Cheng:1992xi,Wise:1992hn},
\begin{eqnarray}
	g_{D^{\ast}D\eta}&=&\alpha \frac{2g}{f_\pi}\sqrt{m_{D^\ast}m_{D}}\ ,\qquad
	g_{D^{\ast}D^\ast \eta }=\alpha \frac{2g}{f_\pi}\ ,\nonumber\\
	g_{ DD\omega }&=&g_{ D^{\ast}D^{\ast} \omega}=\frac{\beta_V g_V}{2}\ ,\nonumber\\
	g_{\omega D^{\ast}D}&=&\frac{\lambda_V g_V}{2}\ ,\quad		f_{\omega D^{\ast}D}=\frac{f_{\omega D^{\ast}D^{\ast}}}{m_{D^*}}=\frac{\lambda_V g_V}{2}\ ,
\end{eqnarray}
where $\beta_V=0.9$, $\lambda_V=0.56$, and $g_V={m_\rho}/{f_\pi}$ with $f_\pi=132\  \mathrm{MeV}$ denoting to the decay constant of pion, and $g=0.59$, which is evaluated by the partial width of $D^\ast\to D\pi$~\cite{ParticleDataGroup:2024cfk}.

\begin{figure}[t]
	\centering
	\includegraphics[width=8.4cm]{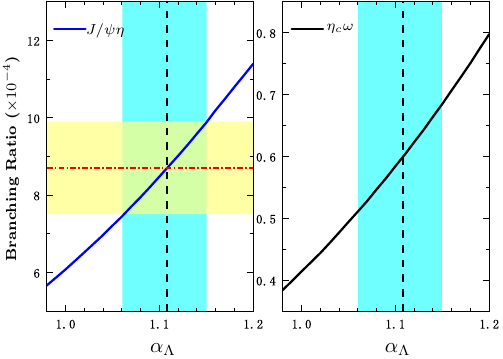}
	\caption{(Color online) The branching fractions of  $\psi(3770) \to J/\psi\eta$ (lift panel) and $\psi(3770) \to \eta_c\omega$ (right panel) depending on the model parameter $\alpha_\Lambda$. The yellow horizontal band with red dashed line is the measured branching fraction of $\psi(3770) \to J/\psi \eta$, and the vertical cyan band with black dashed line indicates the $\alpha_\Lambda$ range determined by comparing our estimations with the experimental measurement. \label{Fig:Br1}}
\end{figure}

\subsection{Decay Widths of The Hidden Charm Decay Processes}
With the above preparations, we can estimate the decay widths of the relevant hidden charm decay processes depending on the model parameter $\alpha_\Lambda$. As we mentioned in the previous section, the parameter $\alpha_\Lambda$ should be of the order of unity, and in the present estimation, we vary the parameter $\alpha_\Lambda$ around one to check the model parameter dependences of the estimated partial widths.
	
The branching fractions of the processes $\psi(3770) \to J/\psi \eta$ and $\psi(3770) \to \eta_c \omega$ depending on the model parameter $\alpha_\Lambda$ are presented in Fig.~\ref{Fig:Br1}. From the figure, one can find both branching fractions increase with the increasing of parameter $\alpha_\Lambda$. In addition, the branching fraction of $\psi(3770) \to J/\psi \eta$ has been measured to be $(8.7\pm 1.2) \times 10^{-4}$~\cite{ParticleDataGroup:2024cfk}, which could be well reproduced in the range of model parameter $\alpha_\Lambda =(1.11^{+0.04}_{-0.05})$. With this fixed model parameter range, we could further estimate the branching fractions of other involved decay processes. For example, the branching fraction of $\psi(3770) \to \eta_c \omega$ is estimated to be $(6.03^{+0.82}_{-0.92})\ \times10^{-5}$, which is about one order smaller than that of $\psi(3770) \to J/\psi \eta$.

\begin{figure}[t]
	\centering
	\includegraphics[width=8.4cm]{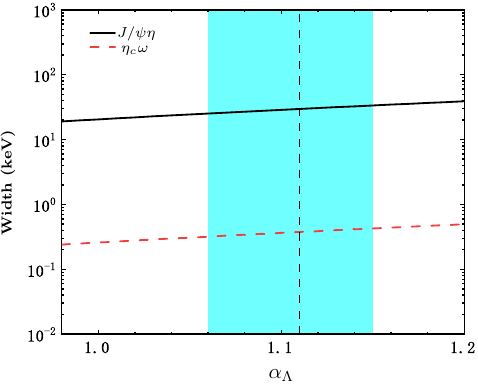}
	\caption{(Color online) The $\alpha_\Lambda$ dependence of the partial widths of $\psi(3823) \to J/\psi \eta$ and $\psi(3823) \to \eta_c \omega$. The vertical cyan band with black dashed lines indicates the parameter range determined by $\psi(3770) \to J/\psi \eta$.  \label{Fig:Br2}}
\end{figure}

Considering the similarity of the of the $D$-wave charmonia, we can investigate the hidden charm decay processes of $\psi_2(3823)$ with the same model parameter range. Currently, the precise determination of the width of $\psi_2(3823)$ remains elusive, with its upper limit $\Gamma_{\psi_2}$ measured to be $2.9$ MeV. Therefore, in the present estimation, we focus on estimating the partial widths of its hidden charm decay processes instead of exploring the branching fractions. The estimated partial widths of $\psi_{2}(3823) \to J/\psi \eta$ and $\psi_2(3823) \to \eta_c \omega$ are presented in Fig. \ref{Fig:Br2}. With the parameter range determined by $\psi(3770) \to J/\psi \eta$, we can obtain,
\begin{eqnarray}
	\Gamma[\psi_2(3823)\to J/\psi\eta]&=&\left(29.64^{+4.01}_{-4.43}\right)\ \mathrm{keV},\nonumber\\
	\Gamma[\psi_2(3823)\to \eta_c \omega]&=&\left(0.38^{+0.05}_{-0.06}\right)\  \mathrm{keV}.
\end{eqnarray}

In addition, the ratio of the partial widths of these two hidden charm decay channels is almost independent on the model parameter $\alpha_\Lambda$, which is estimated to be,
\begin{eqnarray}
R_2=\frac{\Gamma\left[\psi_2(3823)\to \eta_c\omega\right]}{\Gamma\left[\psi_2(3823)\to J/\psi\eta\right]}=0.127,	
\end{eqnarray}	
in the parameter range determined by $\psi(3770)\to J/\psi \eta$.

It should be noted that the BESIII Collaboration searched for the processes $\psi_2(3823) \to \gamma \chi_{c0,1,2},\ \pi^+ \pi^- J/\psi,\ \pi^0 \pi^0 J/\psi,\ \eta J/\psi$ and $\pi^0 J/\psi$ using the reaction $e^+ e^- \to \pi^+ \pi^- \psi_2(3823)$ in a $19\ \mathrm{fb}^{-1}$ data sample collected by the BESIII detector~\cite{BESIII:2021qmo}, the process $\psi_2 \to \gamma \chi_{c1}$ was observed with a higher significance of 11.8$\sigma$, while no significant $\psi_2(3823)$ signals were observed for any other channels. The upper limit of the branching fractions ratio for $\psi_2(3823)\to \eta J/\psi$ relative to $\psi_2(3823) \to \gamma \chi_{c1}$ were reported to be 0.14. In  addition, the Belle Collaboration ever reported the branching fraction of the cascade process $B^{\pm} \to \psi_2(3823) K^{\pm} \to (\chi_{c1} \gamma) K^{\pm} $ to be $(9.7 \pm 2.8 \pm 1.1)\times 10^{-6}$ in 2013~\cite{Belle:2013ewt}, and in 2020, the LHCb Collaboration observed the evidence for $B^+ \to \psi_2(3823) K^+ \to (J/\psi \eta) K^+$ with significance of 3.4 standard deviation, and the  branching fraction of the corresponding cascade process was measured to be  $(1.25^{+0.71}_{-0.53} \pm 0.04)\times 10^{-6}$~\cite{LHCb:2022oqs}, thus, one can conclude the branching fractions ratio for $\psi_2(3823) \to \eta J/\psi$ relative to $\psi_2(3823)\to \gamma \chi_{c1}$ is $(12.9^{+8.3}_{-6.7})\%$, which is consistent with the measurements from the BESIII Collaboration~\cite{BESIII:2021qmo}. In Ref.~\cite{Barnes:2005pb}, the width of $\psi_2(3842) \to \gamma \chi_{c1}$ was estimated to be $268$ and $307$ keV for the non-relativitic and GI relativistic quark models, respectively. Moreover, this width was estimated to be $208\sim 342$ keV in the screened potential model~\cite{Li:2009zu}. Taking the width of $\psi_2(3823) \to \gamma \chi_{c1}$ to be about 300 keV, the present estimations indicate that the branching fractions ratio for $\psi_2(3823)\to J/\psi \eta$ relative to $\psi_2(3823)\to \gamma \chi_{c1}$   is about $10\%$, which is safely under the upper limit reported by the BESIII Collaboration~\cite{BESIII:2021qmo} and consistent with the experimental measurements from Belle and LHCb Collaborations~\cite{Belle:2013ewt, LHCb:2022oqs} .

	\begin{figure}[t]
		\centering
		\includegraphics[width=8.4cm]{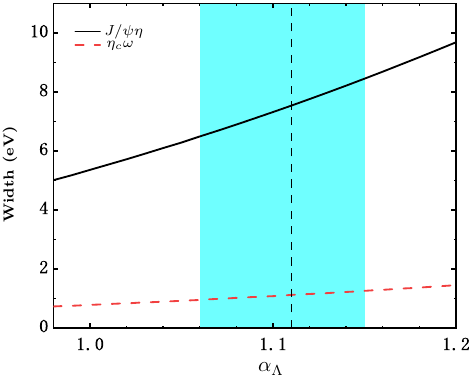}
		\caption{The same as Fig.~\ref{Fig:Br2} but for $\psi_3(3842) \to J/\psi \eta/\eta_c \omega$.\label{Fig:Br3}}
	\end{figure}
	
In a same manner, we can evaluate the hidden charm decays of $\psi_3(3842)$, and the estimated partial decay widths of $\psi_3(3842) \to J/\psi \eta$ and $\eta_c \omega$ are presented in the left panel of Fig.~\ref{Fig:Br3}. In the considered $\alpha_\Lambda$ range determined by $\psi(3770) \to J/\psi \eta$, the partial widths of the hidden charm decay processes of $\psi_3(3842)$ are estimated to be,	
	\begin{eqnarray}
		\Gamma[\psi_3(3842)\to J/\psi\eta]&=&\left(7.54^{+0.91}_{-1.05}\right)\   \mathrm{eV},\nonumber\\
		\Gamma[\psi_3(3842)\to \eta_c \omega]&=&(1.12^{+0.14}_{-0.17})\  \mathrm{eV},
	\end{eqnarray}
which are much small comparing to widths of $\psi_2(3823)$ hidden charm decay processes. It should be noted that the hidden charm decay processes for both $\psi(3770)$ and $\psi_2(3823)$ occur through $P$ wave, while these hidden charm final states couple to $\psi_3(3842)$ via $F$ wave, which should be strongly suppressed. In addition, the coupling $\psi_3(3842) D\bar{D}$ in the meson loops was not considered in the present estimations since this vertex vanish in the leading order of the effective Lagrangian obtained from the expansion of Eq.\eqref{Eq:LagT}, however, the meson loops relevant to $\psi_3(3842) D\bar{D}$ may have sizable contributions since the $D\bar{D}$ in the meson loops could be on-shell.

	\section{Summary}
	\label{Sec:Sum}
The experimental observations of $\psi_2(3823)$ and $\psi_3(3842)$ make $D$ wave charmonia family abundant. Among the spin triplets of the $D$-wave charmonia, $\psi(3770)$ could decay into $D\bar{D}$ through a $P$-wave, however, the experimental measurements indicate that the non-$D\bar{D}$ branching fractions are sizable. Thus, investigating hidden charm decay processes, which is one of the important parts of non-$D\bar{D}$ decay modes, is important for understanding the decay properties of $\psi(3770)$. Moreover, the open charm decay process is forbidden for $\psi_2(3823)$ due to the kinematic limit and $J^P$ quantum numbers conservation, thus, the hidden charm decay processes should be even more important for $\psi_2(3823)$.

In the present work, we investigate the hidden charm decay processes of spin triplets of the $D$-wave charmonia with meson loop mechanism, which is considered to be the phenomenological description of the long-range distributions in the hidden charm decay processes of the higher charmonia. The model parameter $\alpha_\Lambda$ is determined by reproducing the branching fraction of $\psi(3770) \to J/\psi \eta$. With the same model parameter, the branching fraction of $\psi(3770) \to \eta_c \omega$ is predicted to be $(6.03^{+0.82}_{-0.92})\ \times10^{-5}$. In addition, the partial width of $\psi_2(3823) \to J/\psi \eta$ is evaluated to be $\left(29.64^{+4.10}_{-4.43}\right)\ \mathrm{keV}$, and the branching fraction ratio for $\psi_2(3823) \to J/\psi \eta$ relative to $\psi_2(3823) \to \gamma \chi_{c1}$ is about $10\%$, which is safely below the upper limit reported by the BESIII Collaboration and consistent with the experimental measurements from Belle and LHCb Collaborations. More precise measurements from the BESIII, Belle II, and LHCb Collaborations could further test the results in the present work.

\section*{ACKNOWLEDGMENTS}
This work is supported by the National Natural Science Foundation of China under Grant Nos. 12175037, 12335001, and 12405093.	
\begin{widetext}
\appendix	
\section{Amplitudes for $\psi_2(3823)/\psi_3(3842)\to J/\psi \eta$\label{Sec:AppA}}
The amplitudes corresponding to Fig. \ref{Fig:Tri2}-$(g)$-$(k)$ read,	
\begin{eqnarray}
	\mathcal{M}_{g}&=&i^3 \int\frac{d^4 q}{(2\pi)^4}
	\left[g_{\psi_2 D^{*} D} \epsilon^{\mu\nu}(p)(p_{1 \nu}- p_{2 \nu})\right]%\nonumber\\&&
	\left[g_{\eta D^{*} D} p_{3\chi }\right]\left[g_{J / \psi D^{*} D^{*}} \epsilon^{\kappa}\left(p_{4}\right)\left((q_{\xi}-p_{2\xi}) g_{\kappa \phi} +(q_{\phi}-p_{2\phi})g_{\kappa \xi } -(q_{\kappa}-p_{2\kappa})g_{\phi \xi }\right) \right]
	\nonumber\\
	&&\times\frac{1}{p^2_1-m^2_1}\frac{-g_{\mu\phi}+p_2^{\mu} p_2^{\phi} /m^2_2}{p^2_2-m^2_2} \frac{-g_{\chi\xi}+q^{\chi} q^{\xi} /m^2_q}{q^2-m^2_q}\mathcal{F}^2(m_q^2, q^2),\nonumber\\
%\end{eqnarray}
%\begin{eqnarray}
	\mathcal{M}_{h}&=&i^3 \int\frac{d^4 q}{(2\pi)^4}\left[
	g_{\psi_2 D^{*} D} \epsilon^{\mu\nu}(p)(p_{2 \nu}- p_{1 \nu})\right]
	\left[g_{\eta D^{*} D} p_{3\chi }\right]\left[g_{J / \psi D D} \epsilon^{\kappa}\left(p_{4}\right) \left(p_{2\kappa}-q_{\kappa}\right)\right]
	\nonumber\\
	&&\times\frac{-g_{\mu\chi}+p_1^{\mu} p_1^{\chi} /m^2_1}{p^2_1-m^2_1}\frac{1}{p^2_2-m^2_2}\frac{1}{q^2-m^2_q}\mathcal{F}^2(m_q^2,q^2),\nonumber\\
%\end{eqnarray}
%\begin{eqnarray}
	\mathcal{M}_{i}&=&i^3 \int\frac{d^4 q}{(2\pi)^4}\left[
	g_{\psi_2 D^{*} D} \epsilon^{\mu\nu}(p)\left(p_{2 \nu}- p_{1 \nu}\right)\right]
	\left[\frac{1}{2} g_{\eta D^{*} D^{*} } \varepsilon_{\chi\lambda\sigma\rho}  p_{3}^{\lambda} \left(p^{\sigma}_{1}+q^{\sigma}\right)\right] \left[g_{J / \psi D^{*} D} \varepsilon _{\kappa\tau\phi\xi}\left(p^{\xi}_{2}-q^{\xi}\right)  p_{4}^{\kappa } \varepsilon^{\tau}\left(p_{4}\right)\right]\nonumber\\
	&&\times\frac{-g_{\mu\rho}+p_1^{\mu} p_1^{\rho} /m^2_1}{p^2_1-m^2_1}\frac{1}{p^2_2-m^2_2}\frac{-g_{\chi\phi}+q^{\chi} q^{\phi} /m^2_q}{q^2-m^2_q}\mathcal{F}^2(m_q^2,q^2),\nonumber\\
%\end{eqnarray}
%\begin{eqnarray}
	\mathcal{M}_{j}&=&i^3 \int\frac{d^4 q}{(2\pi)^4}
	\Big[ig_{\psi_2 D^{*} D^{*}} \varepsilon_{\alpha\beta\mu\nu}p^{\mu} \epsilon^{\alpha\gamma}(p) \Big((p_{2} ^{\beta}- p_{1}^{\beta})g_{\delta}^{\nu}g_{\theta}^{\gamma}+(p_{2}^{\beta}- p_{1}^{\beta})g_{\gamma\delta}g_{\nu}^{\theta}\Big)\Big] \left[g_{\eta D^{*} D} p_{3\chi }\right]\Big[g_{J / \psi D^{*} D} \varepsilon _{\kappa\tau\phi\xi}\left(q^{\xi}-p^{\xi}_{2}\right)  p_{4}^{\kappa } \epsilon^{\tau}\left(p_{4}\right)\Big]\nonumber\\ &&\times  \frac{1}{q^2-m^2_q}\frac{-g_{\theta\chi}+p_1^{\theta} p_1^{\chi} /m^2_1}{p^2_1-m^2_1} \frac{-g_{\delta\phi}+p_2^{\delta} p_2^{\phi} /m^2_2}{p^2_2-m^2_2}\mathcal{F}^2(m_q^2, q^2),\nonumber\\
%\end{eqnarray}
%\begin{eqnarray}
	\mathcal{M}_{k}&=&i^3 \int\frac{d^4 q}{(2\pi)^4}
	\left[ig_{\psi_2 D^{*} D^{*}} \varepsilon_{\alpha\beta\mu\nu}p^{\mu} \epsilon^{\alpha\gamma}(p) \Big((p_{2} ^{\beta}- p_{1}^{\beta})g_{\delta}^{\nu}g_{\theta}^{\gamma}+(p_{2}^{\beta}- p_{1}^{\beta})g_{\gamma\delta}g_{\nu}^{\theta}\Big)\right]\Big[\frac{1}{2} g_{\eta D^{*} D^{*} } \varepsilon_{\chi\lambda\sigma\rho}  p_{3}^{\lambda} \left(p^{\sigma}_{1}+q^{\sigma}\right)\Big] \left[-g_{J / \psi D^{*} D^{*}}\epsilon^{\kappa}\left(p_{4}\right) \right.\nonumber\\
	&&\times \left.\left((q_{\xi}-p_{2\xi}) g_{\kappa \phi}+(q_{\phi}-p_{2\phi})g_{\kappa \xi } -(q_{\kappa}-p_{2\kappa})g_{\phi \xi }\right) \right] \frac{-g_{\theta\rho}+p_1^{\theta} p_1^{\rho} /m^2_1}{p^2_1-m^2_1}\frac{-g_{\delta\phi}+p_2^{\delta} p_2^{\phi} /m^2_2}{p^2_2-m^2_2} \frac{-g_{\chi\xi}+q^{\chi} q^{\xi} /m^2_q}{q^2-m^2_q}\mathcal{F}^2(m_q^2, q^2).\quad\nonumber\\
\end{eqnarray}
Then the total amplitude for $\psi_2(3823) \to J/\psi \eta$ read,
\begin{eqnarray}
	\mathcal{M}_{\psi_2\to J/\psi \eta} =\mathcal{M}_g +\mathcal{M}_h +\mathcal{M}_i +\mathcal{M}_j +\mathcal{M}_k .
\end{eqnarray}

The amplitudes corresponding to Fig~\ref{Fig:Tri3}-$(l)$-$(m)$ read,
\begin{eqnarray}
	\mathcal{M}_{l}&=&i^3 \int\frac{d^4 q}{(2\pi)^4}
		\Big[ig_{\psi_3 D^{*} D^{*}} \epsilon^{\mu\nu\alpha}(p)\left((p_{2 \mu}- p_{1 \mu})g_{\alpha}^{\theta}g_{\nu}^{\delta}+(p_{2 \mu}- p_{1 \mu})g_{\nu}^{\theta}g_{\alpha}^{\delta}\right)\Big]\Big[g_{\eta D^{*} D} p_{3\chi }\Big]\Big[g_{J / \psi D^{*} D} \varepsilon _{\kappa\tau\phi\xi}\left(q^{\xi}-p^{\xi}_{2}\right) p_{4}^{\kappa } \epsilon^{\tau}\left(p_{4}\right)\Big]\nonumber\\
		&&\times\frac{1}{q^2-m^2_q} \frac{-g_{\theta\chi}+p_1^{\theta} p_1^{\chi} /m^2_1}{p^2_1-m^2_1}\frac{-g_{\delta\phi}+p_2^{\delta} p_2^{\phi} /m^2_2}{p^2_2-m^2_2}\mathcal{F}^2(m_{q}^2, q^2),\nonumber\\
	\mathcal{M}_{m}&=&i^3 \int\frac{d^4 q}{(2\pi)^4}
		\Big[ig_{\psi_3 D^{*} D^{*}} \epsilon^{\mu\nu\alpha}(p)\left((p_{2 \mu}- p_{1 \mu})g_{\alpha}^{\theta}g_{\nu}^{\delta}+(p_{2 \mu}- p_{1 \mu})g_{\nu}^{\theta}g_{\alpha}^{\delta}\right)\Big]\Big[\frac{1}{2} g_{\eta D^{*} D^{*} } \varepsilon_{\chi\lambda\sigma\rho}  p_{3}^{\lambda} \left(p^{\sigma}_{1}+q^{\sigma}\right) \Big]\left[-g_{J / \psi D^{*} D^{*}}\epsilon^{\kappa}\left(p_{4}\right) \right.\nonumber\\
		&&\times\left.\left((q_{\xi}-p_{2\xi}) g_{\kappa \phi}+(q_{\phi}-p_{2\phi})g_{\kappa \xi } -(q_{\kappa}-p_{2\kappa})g_{\phi \xi }\right) \right]
		\frac{-g_{\theta\rho}+p_1^{\theta} p_1^{\rho} /m^2_1}{p^2_1-m^2_1}\frac{-g_{\delta\phi}+p_2^{\delta} p_2^{\phi} /m^2_2}{p^2_2-m^2_2}\times\frac{-g_{\chi\xi}+q^{\chi} q^{\xi} /m^2_q}{q^2-m^2_q}\mathcal{F}^2(m_{q}^2, q^2).\nonumber\\
\end{eqnarray}
Then the total amplitude for $\psi_3(3842) \to J/\psi \eta$ reads
\begin{eqnarray}
	\mathcal{M}_{\psi_3 \to J/\psi \eta} =\mathcal{M}_l +\mathcal{M}_m.
\end{eqnarray}

\begin{figure}[t]
		\begin{tabular}{ccc}
			\centering
			\includegraphics[width=4.2cm]{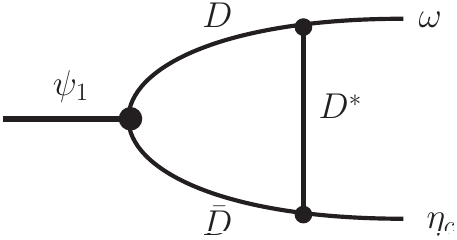}&
			\includegraphics[width=4.2cm]{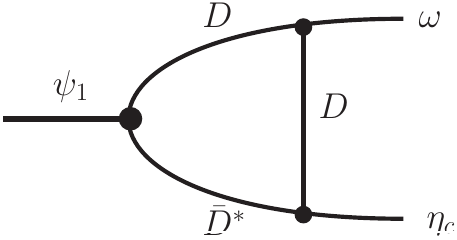}&
			\includegraphics[width=4.2cm]{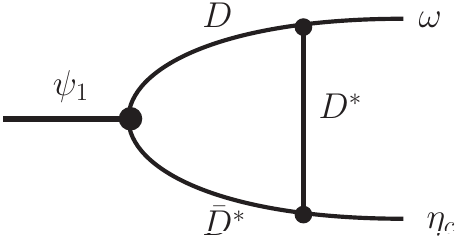}\\ \\
			$(n)$ & $(o)$& $(p)$\\ \\
			\includegraphics[width=4.2cm]{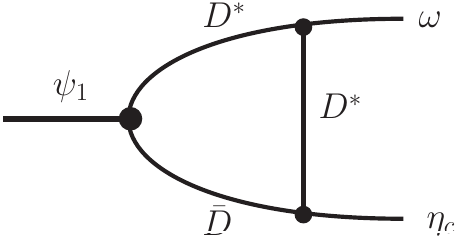}&
			\includegraphics[width=4.2cm]{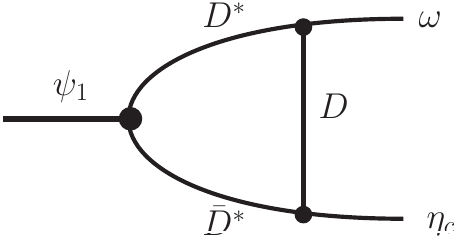}&
			\includegraphics[width=4.2cm]{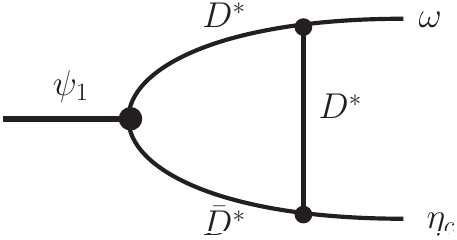}\\ \\
			$(q)$&$(r)$ & $(s)$ 
		\end{tabular}
		\caption{Diagrams contributing to $\psi(3770)\to \eta_c \omega$ at the hadron level .}\label{Fig:Tri4}
	\end{figure}

\section{Amplitudes for $\psi_J(^3D_J)\to \eta_c \omega$\label{Sec:AppB}}
Similar to the processes $\psi(^3D_J)\to J/\psi \eta$, the meson loop contributions to the processes $\psi(^3D_J)\to \eta_c \omega$ are listed in Fig.~\ref{Fig:Tri4}-\ref{Fig:Tri6}. The amplitudes of the process $\psi(3770) \to \eta_c \omega$, corresponding to Fig.~\ref{Fig:Tri4}-$(n)$-$(s)$, reads, 
	
	\begin{eqnarray}
	\mathcal{M}_{n}&=&i^3 \int\frac{d^4 q}{(2\pi)^4}
	\Big[ig_{\psi_1 D D} \varepsilon^{\mu}(p)\left(p_{2 \mu}- p_{1 \mu}\right)\Big]\Big[2f_{\omega D^{*} D} \varepsilon_{\chi\lambda\sigma\rho} p_{3}^{\chi} \varepsilon^{\lambda}\left(p_{3}\right)\left(q^{\sigma}+ p^{\sigma}_{1}\right)\Big] \Big[g_{\eta _{c} D^{*}D}\left(q_{\kappa} - p_{2\kappa}\right)\Big] \nonumber\\
	&&\times\frac{1}{p^2_1-m^2_1}\frac{1}{p^2_2-m^2_2}\frac{-g_{\rho\kappa}+q^{\rho} q^{\kappa} /m^2_q}{q^2-m^2_q}\mathcal{F}^2(m_q^2,q^2),\nonumber\\
	\mathcal{M}_{o}&=&i^3 \int\frac{d^4 q}{(2\pi)^4}
	\Big[g_{\psi_1 D^{*} D} \varepsilon_{\mu \nu \alpha \beta}\left(p^{\mu}_{2}-p^{\mu}_{1}\right)  p^{v}\epsilon^{\alpha}\left(p\right)\Big]\Big[- g_{\omega DD}\left( p_{1\chi}+ q_{\chi}\right) \epsilon^{\chi}\left(p_{3}\right)\Big]\Big[g_{\eta _{c} D^{*}D}\left( q_{\kappa}-p_{2\kappa}\right) \Big]\nonumber\\
	&&\times\frac{1}{p^2_1-m^2_1}\frac{-g_{\beta\kappa}+p_2^{\beta} p_2^{\kappa} /m^2_2}{p^2_2-m^2_2}\frac{1}{q^2-m^2_q}\mathcal{F}^2(m_q^2,q^2),\nonumber\\
    \mathcal{M}_{p}&=&i^3 \int\frac{d^4 q}{(2\pi)^4}
	\Big[g_{\psi_1 D^{*} D} \varepsilon_{\mu \nu \alpha \beta}\left(p^{\mu}_{2}-p^{\mu}_{1}\right)  p^{v}\epsilon^{\alpha}\left(p\right)\Big]\Big[2f_{\omega D^{*} D} \varepsilon_{\chi\lambda\sigma\rho} p_{3}^{\chi} \epsilon^{\lambda}\left(p_{3}\right)\left(q^{\sigma}+ p^{\sigma}_{1}\right)\Big]\Big[g_{\eta_{c} D^{*} D^{*}} \varepsilon_{\kappa\tau\phi\xi}  p_{4}^{\kappa} \left(q^{\phi}- p^{\phi}_{2}\right)\Big] \nonumber\\
	&&\times\frac{1}{p^2_1-m^2_1}\frac{-g_{\beta\xi}+p_2^{\beta} p_2^{\xi} /m^2_2}{p^2_2-m^2_2}\frac{-g_{\rho\tau}+q^{\rho} q^{\tau} /m^2_q}{q^2-m^2_q}\mathcal{F}^2(m_q^2,q^2),\nonumber
\end{eqnarray}
\begin{eqnarray}
	\mathcal{M}_{q}&=&i^3 \int\frac{d^4 q}{(2\pi)^4}\Big[g_{\psi_{1} D^{*} D} \varepsilon_{\mu \nu \alpha \beta} \left(p^{\mu}_{1}-p^{\mu}_{2}\right) p^{\nu} \epsilon^{\alpha}(p)\Big]\Big[g_{\omega D^{*}D^{*}} \left(q_{\sigma}+p_{1\sigma}\right)g^{\rho\lambda}+4f_{\omega D^*D^*} \left(p_{3}^{\lambda}g^{\rho\chi} -p_{3}^{\rho}g_{\sigma}^{\lambda} \right) \Big]\Big[g_{\eta _{c} D^{*}D}\left(q_{\kappa}- p_{2\kappa}\right) \Big]\nonumber\\
	&&\times\frac{-g_{\beta\lambda}+p_1^{\beta} p_1^{\lambda} /m^2_1}{p^2_1-m^2_1}\frac{1}{p^2_2-m^2_2}\frac{-g_{\rho\kappa}+q^{\rho} q^{\kappa} /m^2_q}{q^2-m^2_q}\mathcal{F}^2(m_q^2,q^2),\nonumber\\
    \mathcal{M}_{r}&=&i^3 \int\frac{d^4 q}{(2\pi)^4}\Big[ig_{\psi_{1} D^{*} D^{*}} \epsilon^{\mu}(p)\Big(-4 ( p_{1\mu}-p_{2\mu})  g^{\alpha \beta}+p^{\alpha}_{1\nu}g_{\mu}^{\beta}-p^{\beta}_{2\nu}g_{\alpha}^{\mu}\Big)\Big]\Big[-2 f_{\omega D^{*} D} \varepsilon_{\chi\lambda\sigma\rho} p_{3}^{\chi} \epsilon^{\lambda}\left(p_3\right)\left(q^{\sigma}+p^{\sigma}_{1}\right)\Big]\Big[g_{\eta _{c} D^{*}D}\left(q_{\kappa}-p_{2\kappa}\right) \Big] \nonumber\\
    &&\times\frac{-g_{\beta\rho}+p_1^{\beta} p_1^{\rho} /m^2_1}{p^2_1-m^2_1}\frac{-g_{\alpha\kappa}+p_2^{\alpha} p_2^{\kappa} /m^2_2}{p^2_2-m^2_2}\frac{1}{q^2-m^2_q}\mathcal{F}^2(m_q^2,q^2),\nonumber\\
	\mathcal{M}_{s}&=&i^3\int\frac{d^4 q}{(2\pi)^4}\Big[ig_{\psi_{1} D^{*} D^{*}} \varepsilon^{\mu}(p)\Big(-4 ( p_{1\mu}-p_{2\mu})  g^{\alpha \beta}+p^{\alpha}_{1\nu}g_{\mu}^{\beta}-p^{\beta}_{2\nu}g^{\alpha\mu}\Big)\Big] \left[g_{\omega D^{*}D^{*}} \left(q_{\sigma}+p_{1\sigma}\right)g^{\rho\lambda}+4f_{\omega D^*D^*} \left(p_{3}^{\lambda}g_{\sigma}^{\rho} -p_{3}^{\rho}g_{\sigma}^{\lambda} \right) \right]\nonumber\\
	&& \times\Big[g_{\eta_{c} D^{*} D^{*}} \epsilon_{\kappa\tau\phi\xi}  p_{4}^{\kappa}\left(q^{\phi}- p^{\phi}_{2}\right) \Big] \frac{-g_{\beta\lambda}+p_1^{\beta} p_1^{\lambda} /m^2_1}{p^2_1-m^2_1}\frac{-g_{\alpha\xi}+p_2^{\alpha} p_2^{\xi} /m^2_2}{p^2_2-m^2_2}\frac{-g_{\rho\tau}+q^{\rho} q^{\tau} /m^2_q}{q^2-m^2_q}\mathcal{F}^2(m_q^2,q^2).
\end{eqnarray}
Then the total amplitude for $\psi(3770) \to \eta_c \omega$  reads,
\begin{eqnarray}
\mathcal{M}_{\psi_1\to \eta_c \omega} =\mathcal{M}_n + \mathcal{M}_o +\mathcal{M}_p +\mathcal{M}_q +\mathcal{M}_r +\mathcal{M}_s.	
\end{eqnarray}
	
The amplitudes of $\psi_2(3823) \to \eta_c \omega$ corresponding to Fig.~\ref{Fig:Tri5}-$(t)-(x)$ are,	
	\begin{figure}[t]
		\begin{tabular}{ccc}
			\centering
			\includegraphics[width=4.2cm]{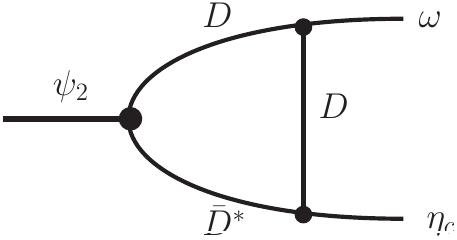}&
			\includegraphics[width=4.2cm]{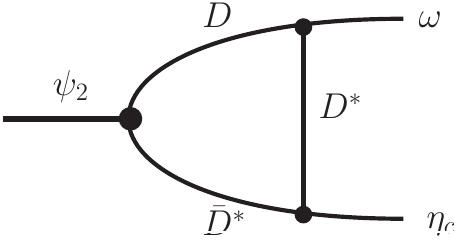}&
			\includegraphics[width=4.2cm]{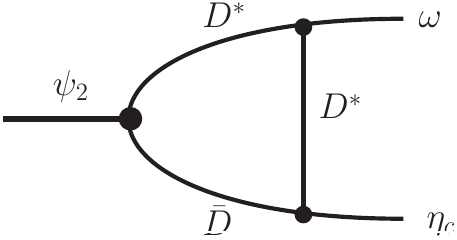}\\ \\
			$(t)$ & $(u)$ &$(v)$ \\ \\
			\includegraphics[width=4.2cm]{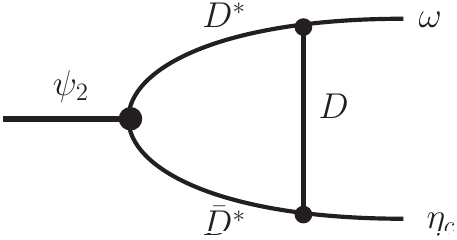}&
			\includegraphics[width=4.2cm]{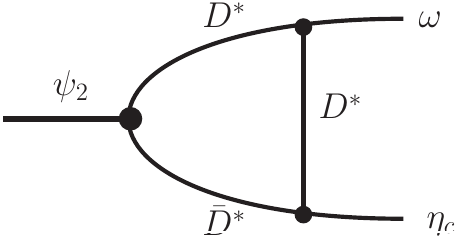}\\ \\
			$(w)$& $(x)$ \\
		\end{tabular}
		\caption{Diagrams contributing to $\psi(3823)\to  \eta_c \omega$ at the hadron level .}\label{Fig:Tri5}
	\end{figure}

	\begin{eqnarray}
	\mathcal{M}_{t}&=&i^3 \int\frac{d^4 q}{(2\pi)^4}
	\Big[g_{\psi_2 D^{*} D} \varepsilon^{\mu\nu}(p)\left(p_{1 \nu}- p_{2 \nu}\right)\Big]\Big[- g_{\omega DD}\left( p_{1\chi}+ q_{\chi}\right) \varepsilon^{\chi}\left(p_{3}\right)\Big] \Big[g_{\eta _{c} D^{*}D}\left( q_{\kappa}- p_{2\kappa}\right) \Big]\nonumber\\
	&& \times\frac{1}{p^2_1-m^2_1} \frac{-g_{\mu\kappa}+p_2^{\mu} p_2^{\kappa} /m^2_2}{p^2_2-m^2_2}\frac{1}{q^2-m^2_q}\mathcal{F}^2(m_q^2,q^2),\nonumber 
    \end{eqnarray}
    \begin{eqnarray}
	\mathcal{M}_{u}&=&i^3 \int\frac{d^4 q}{(2\pi)^4}
	\Big[g_{\psi_2 D^{*} D} \epsilon^{\mu\nu}(p)\left(p_{1 \nu}- p_{2 \nu}\right)\Big]\Big[2f_{\omega D^{*} D} \varepsilon_{\chi\lambda\sigma\rho} p_{3}^{\chi} \epsilon^{\lambda}\left(p_{3}\right)\left(q^{\sigma}+ p^{\sigma}_{1}\right)\Big] \Big[g_{\eta_{c} D^{*} D^{*}} \varepsilon_{\kappa\tau\phi\xi}  p_{4}^{\kappa} \left(q^{\phi}- p^{\phi}_{2}\right) \Big] \nonumber\\
	&&\times\frac{1}{p^2_1-m^2_1}\frac{-g_{\mu\xi}+p_2^{\mu} p_2^{\xi} /m^2_2}{p^2_2-m^2_2}\times\frac{-g_{\rho\tau}+q^{\rho} q^{\tau} /m^2_q}{q^2-m^2_q}\mathcal{F}^2(m_q^2,q^2),\nonumber\\
	\mathcal{M}_{v}&=&i^3 \int\frac{d^4 q}{(2\pi)^4}
	\Big[g_{\psi_2 D^{*} D} \epsilon^{\mu\nu}(p)\left(p_{2 \nu}- p_{1 \nu}\right)\Big]\left[g_{\omega D^{*}D^{*}} \left(q_{\chi}+p_{1\chi}\right)g_{\sigma}^{\chi}g_{\rho}^{\lambda}+4f_{\omega D^*D^*} \left(p_{3}^{\lambda}g_{\rho}^{\chi} -p_{3}^{\chi}g_{\rho}^{\lambda} \right) g_{\chi \rho}\right]\nonumber\\
	&&\times\Big[g_{\eta _{c} D^{*}D}\left(q_{\kappa}- p_{2\kappa}\right)\Big] \frac{-g_{\mu\lambda}+p_1^{\mu} p_1^{\lambda} /m^2_1}{p^2_1-m^2_1}\frac{1}{p^2_2-m^2_2}\frac{-g_{\rho\kappa}+q^{\rho} q^{\kappa} /m^2_q}{q^2-m^2_q}\mathcal{F}^2(m_q^2,q^2),\nonumber\\ 
	\mathcal{M}_{w}&=&i^3 \int\frac{d^4 q}{(2\pi)^4}
	\Big[ig_{\psi_2 D^{*} D^{*}} \varepsilon_{\alpha\beta\mu\nu}p^{\mu} \epsilon^{\alpha\gamma}(p)\left((p_{2} ^{\beta}- p_{1}^{\beta})g_{\delta}^{\nu}g_{\theta}^{\gamma}+(p_{2}^{\beta}- p_{1}^{\beta})g_{\gamma\delta}g_{\nu}^{\theta}\right)\Big]
	\Big[-2 f_{\omega D^{*} D} \varepsilon_{\chi\lambda\sigma\rho} p_{3}^{\chi} \epsilon^{\lambda}\left(p_3\right)\left(q^{\sigma}+p^{\sigma}_{1}\right)\Big]\nonumber\\
	&&\times\Big[g_{\eta _{c} D^{*}D}\left(q_{\kappa}- p_{2\kappa}\right) \Big]\frac{-g_{\theta\rho}+p_1^{\theta} p_1^{\rho} /m^2_1}{p^2_1-m^2_1}\frac{-g_{\delta\kappa}+p_2^{\delta} p_2^{\kappa} /m^2_2}{p^2_2-m^2_2}\frac{1}{q^2-m^2_q}\mathcal{F}^2(m_q^2,q^2),\nonumber\\
	\mathcal{M}_{x}&=&i^3 \int\frac{d^4 q}{(2\pi)^4}
	\Big[ig_{\psi_2 D^{*} D^{*}} \varepsilon_{\alpha\beta\mu\nu}p^{\mu} \epsilon^{\alpha\gamma}(p)\left((p_{2} ^{\beta}- p_{1}^{\beta})g_{\delta}^{\nu}g_{\theta}^{\gamma}+(p_{2}^{\beta}- p_{1}^{\beta})g_{\gamma\delta}g_{\nu}^{\theta}\right)\Big]
	\left[g_{\omega D^{*}D^{*}} \left(q_{\chi}+p_{1\chi}\right)g_{\sigma}^{\chi}g_{\rho}^{\lambda}\right. \nonumber\\
	&&\left.+4f_{\omega D^*D^*} \left(p_{3}^{\lambda}g_{\sigma}^{\chi} -p_{3}^{\chi}g_{\sigma}^{\lambda} \right) g_{\chi \rho}\right]\times\Big[g_{\eta_{c} D^{*} D^{*}} \varepsilon_{\kappa\tau\phi\xi}  p_{4}^{\kappa} \left(q^{\phi}- p^{\phi}_{2}\right) \Big]\frac{-g_{\theta\lambda}+p_1^{\theta} p_1^{\lambda} /m^2_1}{p^2_1-m^2_1}\frac{-g_{\delta\xi}+p_2^{\delta} p_2^{\xi} /m^2_2}{p^2_2-m^2_2}\frac{-g_{\tau\rho}+q^{\tau} q^{\rho} /m^2_q}{q^2-m^2_q}\mathcal{F}^2(m_q^2,q^2).\nonumber\\
\end{eqnarray}
Then the total amplitude for $\psi(3823) \to \eta_c \omega$  reads,
\begin{eqnarray}
	\mathcal{M}_{\psi_2\to \eta_c \omega} =\mathcal{M}_t + \mathcal{M}_u +\mathcal{M}_v +\mathcal{M}_w +\mathcal{M}_x.	
\end{eqnarray}

	\begin{figure}[t]
	\begin{tabular}{ccc}
		\centering
		\includegraphics[width=4.2cm]{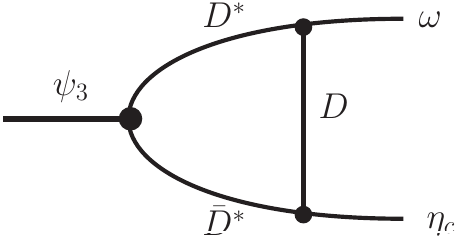}&
		\includegraphics[width=4.2cm]{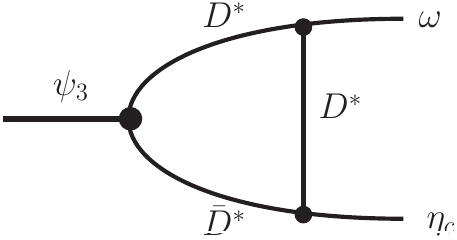}\\
		\\
		$(y)$ & $(z)$  \\		
	\end{tabular}
	\caption{Diagrams contributing to $\psi(3842)\to \eta_c\omega$ at the hadron level. \label{Fig:Tri6}}
\end{figure}

Finally, the ampitudes of $\psi_{3}(3842) \to \eta_c \omega$ corresponding to Fig.~\ref{Fig:Tri6}-$(y)$-$(z)$ read,
	\begin{eqnarray}
		\mathcal{M}_{y}&=&i^3 \int\frac{d^4 q}{(2\pi)^4}
		\Big[ig_{\psi_3 D^{*} D^{*}} \epsilon^{\mu\nu\alpha}(p)\left((p_{2 \mu}- p_{1 \mu})g_{\alpha}^{\theta}g_{\nu}^{\delta}+(p_{2 \mu}- p_{1 \mu})g_{\nu}^{\theta}g_{\alpha}^{\delta}\right)\Big]\Big[-2 f_{\omega D^{*} D} \varepsilon_{\chi\lambda\sigma\rho} p_{3}^{\chi} \left(q^{\sigma}+p^{\sigma}_{1}\right)\Big]\nonumber\\
		&&\times\Big[g_{\eta _{c} D^{*}D}\left(q_{\kappa}- p_{2\kappa}\right)\Big]\frac{-g_{\theta\rho}+p_1^{\theta} p_1^{\rho} /m^2_1}{p^2_1-m^2_1}\frac{-g_{\delta\kappa}+p_2^{\delta} p_2^{\kappa} /m^2_2}{p^2_2-m^2_2}\frac{1}{q^2-m^2_q}\mathcal{F}^2(m_q^2,q^2),\nonumber\\
		\mathcal{M}_{z}&=&i^3 \int\frac{d^4 q}{(2\pi)^4}
		\left[ig_{\psi_3 D^{*} D^{*}} \epsilon^{\mu\nu\alpha}(p)\left((p_{2 \mu}- p_{1 \mu})g_{\alpha}^{\theta}g_{\nu}^{\delta}+(p_{2 \mu}- p_{1 \mu})g_{\nu}^{\theta}g_{\alpha}^{\delta}\right)\right]\left[g_{\omega D^{*}D^{*}} \left(q_{\chi}+p_{1\chi}\right)g_{\sigma}^{\chi}g_{\rho}^{\lambda} +4f_{\omega D^*D^*} \left(p_{3}^{\lambda}g_{\sigma}^{\chi} -p_{3}^{\chi}g_{\sigma}^{\lambda} \right) g_{\chi \rho}\right]\nonumber\\
		&&\times\Big[g_{\eta_{c} D^{*} D^{*}} \epsilon_{\kappa\tau\phi\xi}  p_{4}^{\kappa} \left(q^{\phi}- p^{\phi}_{2}\right) \Big]\frac{-g_{\theta\lambda}+p_1^{\theta} p_1^{\lambda} /m^2_1}{p^2_1-m^2_1}\frac{-g_{\delta\xi}+p_2^{\delta} p_2^{\xi} /m^2_2}{p^2_2-m^2_2}\frac{-g_{\rho\tau}+q^{\rho} q^{\tau} /m^2_q}{q^2-m^2_q}\mathcal{F}^2(m_q^2,q^2).
	\end{eqnarray}
And the total amplitude for $\psi(3842) \to \eta_c \omega$  reads,
	\begin{eqnarray}
		\mathcal{M}_{\psi_3\to \eta_c \omega} =\mathcal{M}_y + \mathcal{M}_z.	
	\end{eqnarray}
	
\end{widetext}
	
%%%%%%%%

\end{document}